\documentclass[%
 reprint,
 amsmath,amssymb,
 aps,
]{revtex4-1}

\usepackage{graphicx}
\usepackage{dcolumn}
\usepackage{bm}
\usepackage{color}
\usepackage{amsmath}

\begin{document}

\preprint{APS/123-QED}

\title{Torsion effects on Condensed Matter: like a magnetic field but not so much}

\author{Anderson A. Lima}
 \email{andersonfisica@hotmail.com}
\affiliation{
Departamento de F\'isica, Universidade Federal da Para\'iba,\\
Caixa Postal 5008, 58051-900, Jo\~ao Pessoa, PB, Brazil 
}
\affiliation{
Centro de Forma\c c\~ao de Professores, Universidade Federal de Campina Grande,\\
58900-000, Cajazeiras, PB, Brazil 
}
\author{Cleverson Filgueiras}
 \homepage{cleverson.filgueiras@dfi.ufla.br}
\affiliation{Departamento de F\'{\i}sica,
           Universidade Federal de Lavras,
           Caixa Postal 3037,
            37200-000 Lavras-MG, Brazil
}
\author{Fernando Moraes}%
 \email{fernando.jsmoraes@ufrpe.br}
\affiliation{
Departamento de F\'isica, Universidade Federal da Para\'iba,\\
Caixa Postal 5008, 58051-900, Jo\~ao Pessoa, PB, Brazil 
} 
\affiliation{Departamento de F\'isica, Universidade Federal Rural de Pernambuco,\\
52171-900, Recife, PE, Brazil}

\date{\today}

\begin{abstract}

In this work, we study the effects of torsion due to a uniform distribution of topological defects (screw dislocations) on free spin/carrier dynamics in elastic solids.  When a particle moves in such a medium, the effect of the torsion associated to  the defect distribution is analogous to that of an applied magnetic field but with subtle differences. Analogue Landau levels are present in this system but they cannot be confined to two dimensions. In the case of spinless carriers, zero modes, which do not appear in the magnetic Landau levels, show up for quantized values of the linear momentum projected on the defects axis. Particles with spin are subjected to a Zeeman-like coupling between  spin and  torsion, which is insensitive to charge. This suggests the possibility of spin resonance experiments without a magnetic field for charged carriers or quasiparticles without electrical charge, like triplet excitons, for instance. 
 
\end{abstract}

\pacs{Valid PACS appear here}
\maketitle


\section{Introduction}

Perhaps because torsion does not play a role as important as curvature in gravitation, it has received little attention from the physics community in general. On the other hand, curvature is ubiquitous: among other things it is the main ingredient in general relativity and recently it appeared as a tool to manipulate electronic properties of low-dimensional systems (see, for instance, Ref. \cite{nano} and references therein). It is the aim of this paper to call the attention to a manifestation of torsion associated to topological defects in electronic/spintronic systems.

The geometric theory of defects of Katanaev and Volovich \cite{Katanaev19921,katanaev} demonstrated the equivalence between three-dimensional gravity with torsion and the theory of defects in solids. In the continuous limit, this approach describes the solid using a Riemann-Cartan manifold where curvature and torsion are associated with disclinations and dislocations, respectively. The Burgers vector of the dislocation is associated with torsion and the Frank angle of the disclination is associated with curvature. In this theory, the elastic deformations introduced by defects in the medium are incorporated into the metric of the manifold. This way,   geometric tools from general relativity can then be used in the study  of physical properties of material media containing topological defects. Conversely, these media can provide experimental tests of hypotheses of gravitational theory involving topological defects \cite{MORAES2000}.

Topological defects are present in any type of crystalline material. They are a result of a local break of the discrete symmetry of crystals. Their generation as a ``cut and glue'' process   was first described  by Vito Volterra in 1907 \cite{Volterra1907} in what became known as the Volterra process. The geometrical description of elastic media with topological defects started in the 1950's with Kondo \cite{kondo1958memoirs} and Bilby, Bullough and Smith  \cite{Bilby} who, independently, build an elegant differential geometric theory of dislocations. The continuum theory of defects had then very important contributions in the ensuing years from Eshelby \cite{eshelby1956continuum}, Kr\"oner \cite{kroner1956kontinuumstheorie} and Zorawski \cite{zorawski1967theorie}. A very nice review of these developments is found in \cite{kroner1981continuum}, including dislocations, disclinations and point defects. Dislocations, in particular,  have a strong influence on the elastic and electronic properties of the media where they occur. Scattering by these defects has strong effects on elastic waves \cite{maurel2008interaction}, magnons \cite{woltersdorf2004two} and electrons \citep{Jaszek2001}.  For the electronic device industry, these defects are a problem since they interfere in the electronic  properties of the materials by way of scattering or  acting as electronic traps.  On the other hand, line defects might be useful in the design of electronic devices as they act as one-dimensional conductive channels \cite{vish,jairo, knut}. There is large evidence of the similarity between dislocation lines and magnetic flux tubes. Recent research on the effects of dislocations indicate that they act as sources of a ficticious magnetic field in a Weyl semimetal \cite{Sumiyoshi}. On the other hand, even though they cause effects like Aharonov-Bohm-like phases \cite{Kawamura1978,PhysRevB.59.13491,Furtado2001160} and induce transverse modes in the quantum Hall effect \cite{clev}, it has been shown recently, in a study of optical conductivity in the presence of a screw dislocation  \cite{Taira}, that they act qualitatively different from the magnetic field. Therefore,  research on defects and how they may influence the dynamics of  carriers is important for the improvement of electronic  technology, the discovery of new phenomena and better control of  transmission processes. The advent of spintronics \cite{RevModPhys.76.323}, rises the interest on the effects of topological defects on the dynamics of the spinning carriers. Since torsion couples to spin, understanding its effects on spin dynamics may help in the design of spintronic devices through dislocation engineering.

Using the geometric theory of defects, the authors of Ref. \cite{elastic} studied the quantum dynamics of  a particle moving in a screw dislocation density, homogeneously distributed, and parallel to each other. Even without the application of an external magnetic field, they obtained energy eigenvalues much similar to the well known Landau levels due to the torsion field associated to the defect distribution. The coupling of torsion to the wave function comes from the distortion of the medium with the defects, which effectively appears as a modification of the kinetic energy due to the distorted trajectories. The Laplacian operator is substituted by its more general version, the Laplace-Beltrami operator, which incorporates the effective geometry of the medium as seen from the point of view of the geometric theory of defects. In the present work, we go a step further and include the coupling between the torsion field and the  magnetic moment of the particle, a well known phenomenon of magnetoelasticity \cite{Eringen}. In elastic materials, the distortion of the medium around dislocations tends to align  magnetic moments creating nanoscale magnetic patterns which have been  largely studied both from the theoretical \cite{Seeger} as well from the experimental \cite{Bode} points of view. Here, we present a simplified approach, inspired in the geometric theory of defects, by adding to the Hamiltonian studied in Ref. \cite{elastic}, a term coupling the torsion field of the defects to the spin (and therefore to the magnetic moment) of the quantum particle. This coupling is well known in  quantum field theory in the presence of torsion \cite{shapiro} and naturally does not include all the details of the magnetoelasticity theory but, nevertheless, provides a geometric framework for the study of spinning particles in defected media.   

This work is presented as follows:  in Section \ref{sect2}, after a brief introduction (subsection \ref{dis}) to screw dislocations and their geometrical description, we  discuss the classical behavior (Subsection \ref{cla}) and solve (Subsection \ref{qua}) the Schr\"odinger-Pauli-like equation for a free  particle with spin interacting with an homogeneous distribution of screw dislocations. In the final part of this subsection we discuss the results and their implications to real systems like GaAs, for instance. In Section \ref{conc} we present our conclusions. Appendix \ref{back} gives some background on the geometry involving torsion   and Appendix \ref{tqm} gives the derivation of the torsion version of the Schr\"odinger-Pauli equation  used in Subsection \ref{qua}. 

\section{Particle dynamics in  a uniform screw dislocation distribution \label{sect2}}

\subsection{Screw dislocations \label{dis}}

\begin{figure}[!htb]
\begin{center}
\includegraphics[width=5cm]{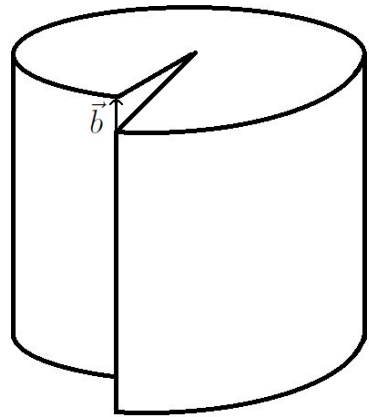}
\caption{Representation of a continuous elastic medium with a single screw dislocation with its Burgers vector $\vec{b}$.}
\label{fig1}
\end{center}
\end{figure}

The discrete translational symmetry of crystals eventually fails in real materials. A quite common fault that appears both in metals and semiconductors  is the screw dislocation.   Such defects appear naturally in the fabrication processes of the materials but are usually eliminated by annealing \cite{rollett2004recrystallization}, since they impair  the performance of electronic devices, as mentioned in the Introduction. A single screw dislocation can be visualized  by the Volterra process \cite{Volterra1907}: make a cut  in the material and move the parts  separated by the cut relative to each other, as shown in Fig. \ref{fig1}. The displacement made is known as the Burgers vector, represented by $\vec{b}$ in the figure. In real crystals, the displacement must be a translation vector of the lattice so that the cut surfaces fit together perfectly, except near the defect axis, defined by the end of the cut.

It is then clear that, in a medium with a screw dislocation with Burgers vector $\vec{b}=b\hat{z}$,  given the cylindrical coordinates $(\rho,\varphi ,z)$, as $\varphi \rightarrow \varphi + 2\pi$ we must have  $z \rightarrow z + b$. These are boundary conditions which is naturally encoded in a geometric background in the framework of the geometric theory of defects \cite{Katanaev19921,katanaev}. For this particular case of a single screw dislocation, the corresponding effective geometry is described by the line element \cite{Furtado2001160}
\begin{equation}
ds^2= d\rho^2 + \rho^2 d\varphi^2 + \left( dz + \frac{b}{2\pi} d\varphi \right)^2 .
\label{metricsingle}
\end{equation}
The generalization of Eq. (\ref{metricsingle}) to a continuous distribution of screw dislocations is \cite{elastic}
\begin{equation}
ds^2= d\rho^2 + \rho^2 d\varphi^2 + \left( dz + \Omega \rho ^2 d\varphi \right)^2,
\label{metricdens}
\end{equation}
where the density of the Burgers vectors $\Omega = bN/ 2$ and $N$ is the surface density of dislocations. Eq.  (\ref{metricdens})  sets the background for our classical and quantum studies that follow.


\subsection{Classical behavior \label{cla}}
 
In order to gain some intuition on the dynamics of a particle in the presence of a screw dislocation density we describe initially its classical behavior. 
Starting with the line element (\ref{metricdens}) one can get the squared velocity of a particle of mass $\mu$ and, consequently, the Lagrangian
\begin{equation}
L=\frac{1}{2} \mu \left(\frac{ds}{dt}\right)^2 = \frac{1}{2} \mu \left(\dot{\rho}^2 + \rho^2 \dot{\varphi}^2 + (\dot{z} + \Omega \rho^2 \dot{\varphi})^2 \right). \label{Lag}
\end{equation}
In contrast, the analogous Lagrangian for a particle of charge $q$ in an uniform magnetic field $\vec{B}=B\hat{z}$ is  \cite{symon1971mechanics}  
\begin{equation}
L^{mag}= \frac{1}{2} \mu \left(\dot{\rho}^2 + \rho^2 \dot{\varphi}^2  +\dot{z}^2 \right)+ \frac{1}{2}q \rho^2 \dot{\varphi}B . \label{Lagmag}
\end{equation}
It is clear that  $z$  and $\varphi$  are cyclic coordinates for the Lagrangian (\ref{Lag}) and therefore that the corresponding momenta are conserved. That is,
\begin{equation}
p_z = \frac{\partial L}{\partial \dot{z}}=  \mu \dot{z} + \mu\Omega \rho^2 \dot{\varphi}  \label{pz}
\end{equation}
and
\begin{equation}
p_{\varphi} = \frac{\partial L}{\partial \dot{\varphi}}= \mu\rho^2 \dot{\varphi} + \Omega \rho^2  p_z ,\label{el}
\end{equation}
where $p_{\varphi}$ and $p_z$ are constants of motion.  The corresponding momenta for the magnetic case are \cite{symon1971mechanics}   
\begin{equation}
p_z^{mag} = \mu \dot{z} \label{pzmag}
\end{equation}
and 
\begin{equation}
p_{\varphi}^{mag}=\mu\rho^2 \dot{\varphi} + \frac{1}{2}q\rho^2 B. \label{elmag}
\end{equation}

\begin{figure}[!htb]
\begin{center}
\includegraphics[width=6cm,width=6cm]{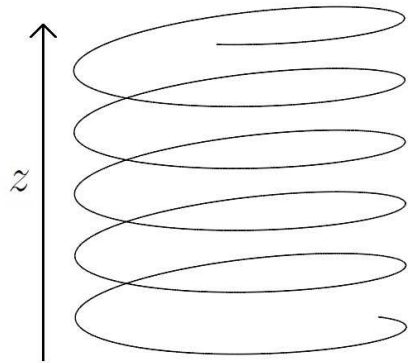}
\caption{Classical trajectory for a particle moving in the presence of a continuous distribution of screw dislocations.}
\label{geo}
\end{center}
\end{figure}

The third constant of motion, the energy $E$, is given by the Hamiltonian $H=\sum_i p_i \dot{q}_i - L(q_i , \dot{q}_i )$, which becomes
\begin{equation}
H = \frac{1}{2}\mu\dot{\rho}^2 + \frac{1}{2}\mu\dot{z}^2 + \frac{1}{2}\mu\left( 1 + \Omega^2 \rho^2 \right)\rho^2\dot{\varphi}^2 + \mu\Omega \rho^2 \dot{\varphi}\dot{z}
\end{equation}
and therefore
\begin{equation}
E= \frac{1}{2}\mu\dot{\rho}^2  +  \frac{p_z^2}{2\mu} +  \frac{1}{2\mu} \left( \frac{p_{\varphi}}{\rho} - \Omega\rho p_z \right)^2 , \label{Losc}
\end{equation}
after using  Eqs.(\ref{pz}) and (\ref{el}).  The corresponding expression for the magnetic case has exactly the same form \cite{symon1971mechanics}: 
\begin{equation}
 E^{mag}= \frac{1}{2}\mu\dot{\rho}^2  +   \frac{(p_z^{mag})^2}{2\mu}   +  \frac{1}{2\mu} \left( \frac{p_{\varphi}^{mag}}{\rho} - \frac{1}{2}q\rho B\right)^2 . \label{Loscmag}
 \end{equation}
The steady orbits are obtained at the minimum of the effective potential $\frac{1}{2\mu} \left( \frac{p_{\varphi}}{\rho} - \Omega\rho p_z \right)^2 $. This requirement, plus  Eqs. (\ref{pz}) and (\ref{el}) gives for the angular frequency,
\begin{equation}
\dot{\varphi}=- \frac{2\Omega p_z}{\mu}, \label{angfreq}
\end{equation}
and for the radius
\begin{equation}
\rho=\sqrt{\frac{|p_{\varphi}|}{\Omega |p_z |}}, \label{raio}
\end{equation}
to be compared with the magnetic case
\begin{equation}
\dot{\varphi}^{mag}=- \frac{qB}{\mu}, \label{angfreqmag}
\end{equation}
\begin{equation}
\rho^{mag}=\sqrt{\frac{2|p_{\varphi}|}{q|B|}}, \label{raiomag}
\end{equation}
respectively. Note that $p_{\varphi}=-\Omega \rho^2 p_z$, and therefore that, $p_z$ and $p_{\varphi}$ have opposing signs. Clearly the orbits are spirals like the one shown in Fig. \ref{geo}, as in the magnetic case. 

Inspection of the above equations indicate that $2\Omega p_z$ is the analogue of $qB$ in the defect distribution. This indicates that absence of motion along the $z$ direction is equivalent to no magnetic field. Planar, circular orbits are not possible in the defect distribution, in contrast with the cyclotron orbits of the charged particle moving in a magnetic field. Furthermore,  while the magnetic field couples to the charge, the defect density couples to the mass of the particle by modifying its inertial properties. This way, a classical Hall-like effect due to a screw dislocation density is unable to determine the charge of a carrier.

\begin{figure}[!htb]
\begin{center}
\includegraphics[width=8cm,width=8cm]{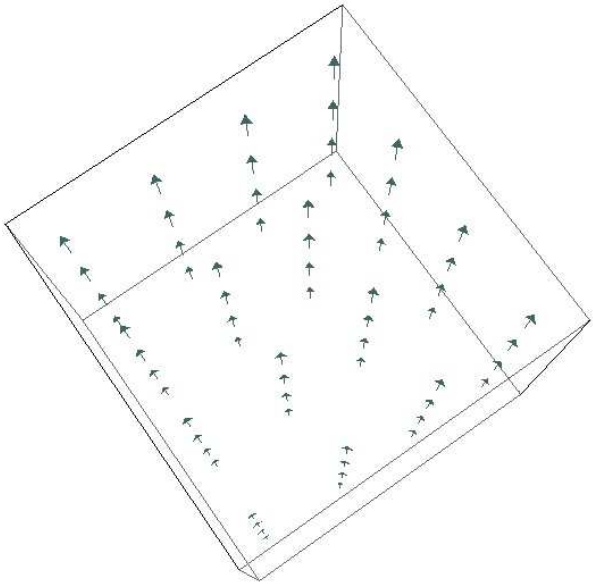}
\caption{The torsion field associated to the screw dislocation distribution, as represented by the axial vector  (Eq. (\ref{S3})). The field is oriented parallel to the $z$ direction. } 
\label{graf}
\end{center}
\end{figure}

The problem considered here involves a uniform torsion field, which is represented by its axial vector (Eq. (\ref{S3})), as described in Appendix \ref{back}. A representation of this vector field is shown in Fig. \ref{graf} but a more intuitive picture of its effect can be obtained by exploring the isometries associated to this field.
According to the  well-known Noether theorem, conservation laws are associated to symmetry operations that leave the Lagrangian invariant. The conserved quantities are generators of these symmetry operations. Thus, the conserved mixed momenta given by Eqs. (\ref{pz})  and (\ref{el}) generate a combined motion along the $z$ and $\varphi$ directions. In other words, while the momentum $p_{\varphi}$  generates  infinitesimal rotations around the $z$ axis, the momentum $p_z$ generates  simultaneous displacements along $z$, due to the coupling between Eqs. (\ref{pz})  and (\ref{el}).  

\begin{figure}[!htb]
\begin{center}
\includegraphics[width=8cm,width=8cm]{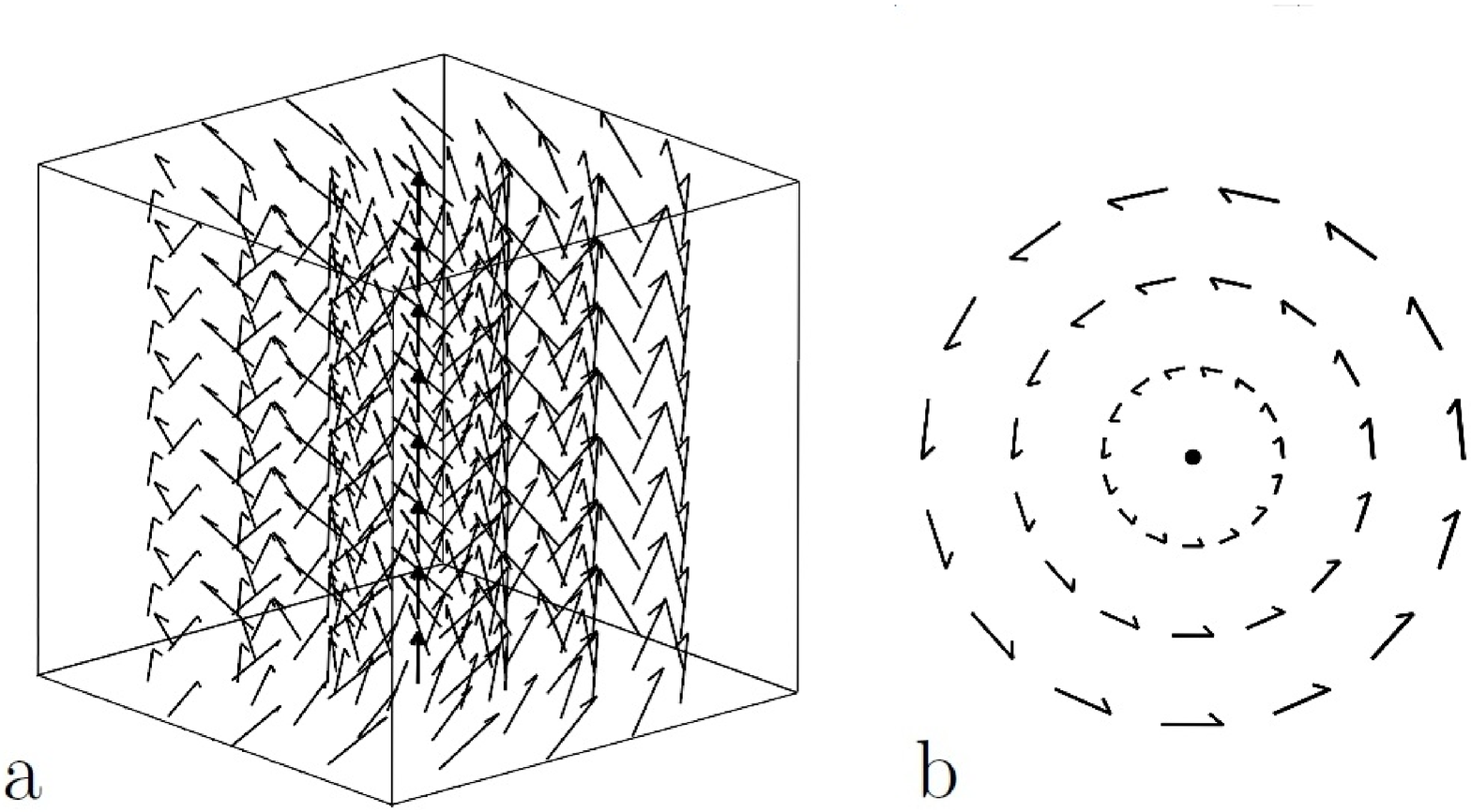}
\caption{Representation of the vector field $ (\hat{z} + \Omega \rho \hat{\varphi})$ which gives, at each point, the direction of an infinitesimal translation that leaves the equations of motion invariant: a) side view and b) top view.} 
\label{graf2}
\end{center}
\end{figure}

From Eqs. (\ref{pz}) and (\ref{el}), we see that during an infinitesimal interval of time $dt$ there will be simultaneous translations of  $dz$ along the $\hat{z}$ direction and $\Omega\rho (\rho d\varphi)$ along the $\hat{\varphi}$ direction which leave  the Lagrangian invariant. In order to give an intuitive view of the isometries  introduced by the torsion field we show in Fig. \ref{graf2} a representation of the vector field $ (\hat{z} + \Omega \rho \hat{\varphi})$ which represents the infinitesimal displacements just mentioned.

\subsection{Quantum behavior \label{qua}}

Now, we turn to the quantum behavior. The dynamics of the particle in the screw dislocation distribution is given by  an expression analogous to the Schr\"odinger-Pauli Hamiltonian, which was derived in Appendix \ref{tqm} (Eq. (\ref{Htor})): 
\begin{equation}
H= \frac{1}{2\mu}\vec{p}\,^2  + \eta \vec{\sigma} \cdot \vec{S} , \label{Htor1}
\end{equation}
where $\mu$ is the effective mass of the particle in the medium with the defects, $\vec{\sigma}$ is the particle's spin and $\vec{S}$ is the torsion field associated to the defect distribution.  The second term on the right hand side of this equation is analogous to the Zeeman term $\left( \frac{e\hbar}{2m}\right)\vec{\sigma} \cdot \vec {B}$ which couples spin angular momentum and magnetic field.  The material-dependent coupling constant $\eta$ expresses the magnetoelastic interaction; that is, it describes how strong is the effect of elastic torsion, due to the defect distribution, on the spin angular momentum of the particle and should be determined experimentally for each material. Note that $\eta$ has the dimensions of energy times distance.  Shapiro and co-authors have shown that $\eta=-(1/8)\hbar c$ and they also consider the possible generalization to other situations \cite{shapiro2,Ryder199821}. Using the same  value, we obtain spin splitting of order of magnitude between  those observed experimentally  \cite{kato2004coherent} and theoretically predicted  \cite{chantis2008strain} for strained GaAs, as shown below. 

Eq. (\ref{Htor1}) leads to the following differential equation in cylindrical coordinates (see the derivation of Eq. (\ref{schrod}) in  Appendix \ref{tqm}):
\begin{eqnarray}
&&\left[ \frac{1}{\rho}\frac{\partial}{\partial \rho} \left( \rho \frac{\partial}{\partial \rho} \right) + \frac{1}{\rho^2}\frac{\partial^2}{\partial \varphi^2} - 2\Omega \frac{\partial^2}{\partial z \partial \varphi} \right. + \nonumber \\
&& \left. \left( 1+\Omega^2 \rho^2 \right)\frac{\partial^2}{\partial z^2}+\frac{2\mu E}{\hbar^2}-\frac{4\mu \Omega \eta \lambda}{\hbar^2} \right] \Psi = 0 ,
\label{schrod1}
\end{eqnarray}
where $\lambda=\pm 1$ is the eigenvalue of the Pauli matrix $\sigma^3$.
To solve this equation we use the  {\it ansatz}
\begin{equation}
\Psi \left( \rho, \varphi, z \right) = R \left( \rho \right) e^{im \varphi}e^{ikz} ,
\label{ansatz}
\end{equation}
where $m$ is an integer. Substituting Eq.(\ref{ansatz}) into Eq.(\ref{schrod1}), the Scr\" odinger-Pauli-like equation takes the following form
\begin{eqnarray}
&&\frac{1}{\rho}\frac{d}{d\rho}\left( \rho \frac{dR}{d\rho}\right)-\frac{m^2}{\rho^2}R - k^2 \Omega^2 \rho^2 R + \nonumber \\
&& \left[ \frac{2\mu E}{\hbar^2} +2mk\Omega - k^2 -\frac{4\mu \Omega \eta \lambda}{\hbar^2}  \right] R =0 .
\label{xx}
\end{eqnarray}
Comparing this equation to its classical counterpart, Eq. (\ref{Losc}), we identify that the effective classical potential $ \frac{1}{2\mu} \left( \frac{p_{\varphi}}{\rho} - \Omega\rho p_z \right)^2$ is mapped onto $\frac{1}{2\mu} \left( \frac{\hbar m}{\rho} - \Omega\rho \hbar k \right)^2$ as it should.

Following Ref. \citep{elastic}, we consider the change of variables
\begin{equation}
\xi \equiv k \Omega \rho^2 ,
\label{changev}
\end{equation}
and substituting  Eq.(\ref{changev}) into Eq.(\ref{xx}), we have the following equation
\begin{equation}
\xi \frac{d^2 R}{d\xi^2}+\frac{dR}{d\xi}-\frac{m^2}{4\xi}R -\frac{\xi}{4}R+\beta R =0 ,
\label{R}
\end{equation}
where 

\begin{equation}
\beta = \frac{1}{4k\Omega} \left( \frac{2\mu E}{\hbar^2} +2mk\Omega - k^2 -\frac{4\mu \Omega \eta \lambda}{\hbar^2}  \right). \label{beta}
\end{equation}

The asymptotic behavior of Eq. (\ref{R}) suggests that we write
\begin{equation}
R \left( \xi \right) = e^{-\frac{\xi}{2}}\xi^{\frac{|m|}{2}}u\left( \xi \right) .
\label{}
\end{equation}
The equation for $u(\xi)$ is therefore
\begin{equation}
\xi \frac{d^2 u}{d\xi^2}+\left( 1+|m|-\xi \right)\frac{du}{d\xi} + \left( \beta -\frac{|m|+1}{2}\right) u=0 .
\label{function}
\end{equation}

This second order ordinary  differential equation, which has  as coefficients  linear functions of $ \xi $, has as solution a confluent hypergeometric function, given by
\begin{equation}
u\left( \xi \right) = F\left( -\beta + \frac{|m|+1}{2},1+|m|,\xi \right).
\label{n}
\end{equation}

For a normalisable wave function, the series (\ref{n}) should end, becoming a polynomial of degree $n_{\rho}$ so that
\begin{equation}
-\beta + \frac{|m|+1}{2}=-n_{\rho}
\label{nn}
\end{equation}

\begin{figure}[!htb]
\begin{center}
\includegraphics[width=9cm,width=9cm]{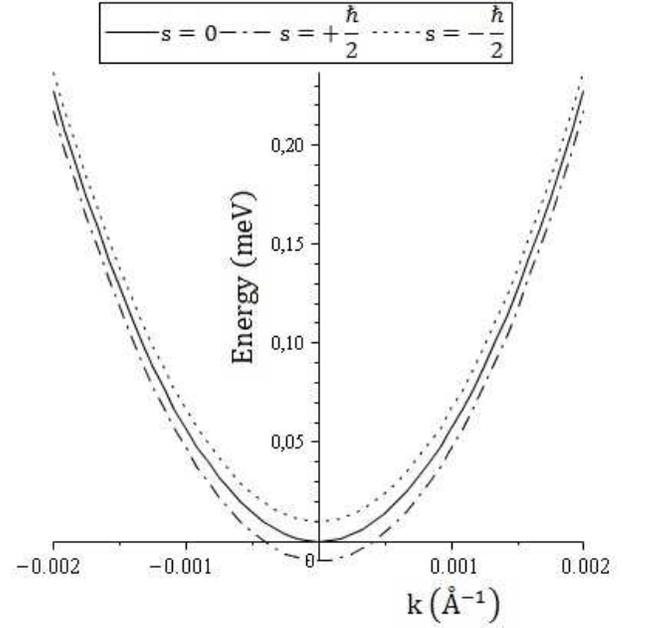}
\caption{The energy, as in Eq. (\ref{ll}), as a function of the wave vector for a dislocation density $N=10^8$ disl.$/cm^2$.  We consider $s=0$ for the particle without spin and $s= \pm \frac{\hbar}{2}$ for the case with spin.  We used $\eta = -\frac{1}{8}\hbar c$ and data for GaAS (electron effective mass, Burgers vector magnitude, and typical range of dislocation density). } 
\label{grafico}
\end{center}
\end{figure}

Combining  Eqs. (\ref{nn}) and (\ref{beta}), we find the discrete energy values with the spin term contribution
\begin{equation}
E=\hbar \omega_{el} \left(n+\frac{1}{2} \right) + 2\Omega \eta \lambda + \frac{k^2 \hbar^2}{2 \mu},
\label{ll}
\end{equation}
where 
\begin{equation}
\omega_ {el} = \frac{2 \hbar k \Omega}{\mu} \label{freq}
\end{equation}
is the elastic ``cyclotron''  frequency and $n=n_{\rho} + \frac{|m|}{2}+\frac{m}{2}$. Note that Eq. (\ref{freq}) is consistent with its classical counterpart (\ref{angfreq}) with the identification of $\hbar k$ with $p_z$. Note also that the quantum numbers $n_{\rho}$, $m$ and $k$ correspond to the radial, angular and linear degrees of freedom, respectively. The Burgers vector density, as previously mentioned, is $\Omega = bN/2$, where $b$ is the Burgers vector and $N$ is the surface density of screw dislocations. Like in the case of magnetic Landau levels,  $E$ is degenerate with respect to $n$, since $n_{\rho} \in \mathbb{N}$  and $m \in \mathbb{Z}$. The spin term partially removes the degeneracy as seen in Fig. \ref{grafico}. The fact that the elastic frequency (\ref{freq}) depends on $k$ expresses the 3-dimensionality of the motion. Since the defects couple $z$ and $\varphi$, their effect does not appear without motion along the $z$-direction, as mentioned in Subection \ref{cla}. The fact that $k \in \mathbb{R}$ brings about the possibility of zero modes for spinless particles,  when $k =- (2n+1) 2\Omega = -(2n+1) S$ and $k=0$. 


Fig. \ref{grafico} shows a plot of the energy as a function of wave vector $k$ showing the vertical shift of the parabolic band due to spin splitting. The aforementioned zero mode at $k =- (2n+1) 2\Omega = -(2n+1) S$ corresponds to a value of $k$ too small to be seen in the scale used. 

\begin{figure}[!htb]
\begin{center}
\includegraphics[width=8cm,width=8cm]{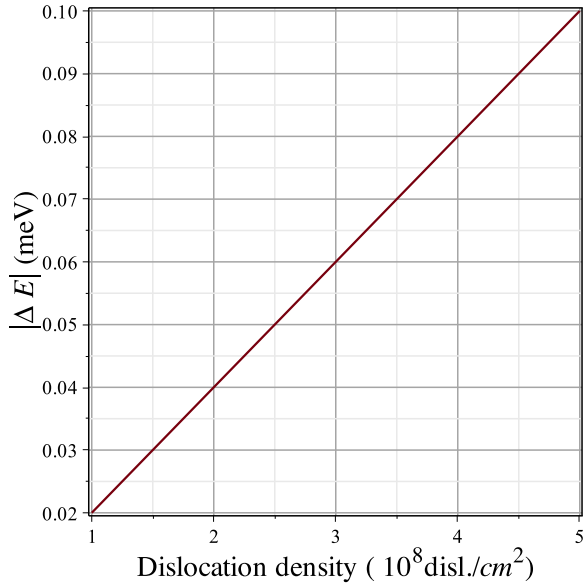}
\caption{The magnitude of the spin splitting, as a function of  dislocation density.   } 
\label{grafico2}
\end{center}
\end{figure}

From Eq. (\ref{ll}) we have the magnitude of the spin splitting, $|\Delta E| = 4 \Omega|\eta|$, which is plotted in Fig. \ref{grafico2} as a function of the dislocation density. We remark that, from the experimental point of view, this   would require separate samples with different defect densities. In order to plot the graphics shown in Fig. \ref{grafico} and \ref{grafico2}, we use parameters for bulk GaAs found in the literature \cite{Jaszek2001}: the density of screw dislocations $N \sim 10^8$ disl./$cm^2$), the magnitude of the Burgers vector $b=4 \times 10^{-10} m$ , and the electron effective mass is $\mu = 0.61 \times 10^{-31}kg$.

What is seen in Fig. \ref{grafico2} is the same kind of splitting that happens in the Zeeman effect, which leads to Electron Spin Resonance (ESR) phenomena \cite{esr}, suggesting the possibility of ESR experiments using elastic torsion instead of magnetic field. Furthermore, as pointed out in \cite{kato2004coherent} and \cite{li2008strain}, strain can be used to manipulate spins, turning topological defects like dislocations, with their associated strain/torsion field, into possible tools for this manipulation.

The results obtained here can be compared to those of references \cite{kato2004coherent} and \cite{chantis2008strain}, for instance, that study spin-splitting in GaAs strained samples. The relevant parameters for spin splitting in these works are, respectively, drift velocity \cite{kato2004coherent} and wave vector amplitude \cite{chantis2008strain}. In our work, the relevant parameter is the dislocation density, that is, the strain itself. For the range of drift velocities studied experimentally in Ref. \cite{kato2004coherent}, the authors report $|\Delta E| \sim 0.1\, \mu$eV. For the range of wave vector amplitudes studied theoretically in Ref.\cite{chantis2008strain},  it is $|\Delta E| \sim 0.1$ meV. Our results, for realistic dislocation densities, are placed in between, with $|\Delta E| \sim 0.01$ meV, making their  experimental observation feasible.

\section{Concluding remarks \label{conc}}

In this work, we study the dynamics of a particle in a medium with a uniform distribution of screw dislocations, which are quite common defects in crystalline materials.  Although there are some similarities with the problem of a charged particle in the presence of a uniform magnetic field, as it has been pointed out in the literature, there are  important differences as well. The classical picture, analyzed in Subsection \ref{cla} as a means of providing some physical intuition on the effects of the  defect distribution, presents  some of these differences. For instance,  the motion is always three-dimensional due to the nature of the defect distribution, which couples the angular ($\varphi$) and the linear ($z$) degrees of freedom of the particle. Also, there is no distinction between positively or negatively charged particles.  

The quantum picture introduces a coupling between the spin of the particle and the defect distribution. It is known that such a defect density on elastic media generates a torsion field that acts on  the particle as if an  external magnetic field were being applied to it.  In fact, a Zeeman-like effect appears due to the torsion-spin coupling. On the other hand, the energy levels, although similar to the magnetic (Landau) levels, have a few important differences. While the motion of charged particles in the presence of a uniform magnetic field may be confined to a plane perpendicular to the field lines, this is not possible in the case of torsion, which needs motion in three-dimensional space in order to show up its effects. Furthermore, in the case of a spinless particle, when  the $z$-component of its linear momentum is antiparallel to the torsion field, there appear zero energy modes for  quantized values of the momentum. 

The coupling of the torsion field to the spin of the particle introduces a shift in the energy spectrum and breaks the spin degeneracy  like in the Zeeman effect. This  magnetic-like field produced by the  density of defects is insensitive to the  charge signal and  thus will have the same effect on spinning neutral particles. Since the Zeeman-like splitting allows for ESR experiments, there is the perspective of  using  magnetic-like resonance to study neutral particles like excitons in the triplet state. Even though the main focus of spintronic studies has been on dislocation-free materials, since the defects couple to magnetic moments,  they may be useful in the design of spintronic devices, in addition to curvature, which has been suggested as a tool to aid in the design of electronic nanodevices \cite{nano}.

\appendix

\section{Some background on torsion \label{back}}

Riemanniann (and pseudo-Riemmanian as in general relativity) geometry is torsion-free due to the requirement that the metric tensor should be symmetric:
\begin{equation}
g_{\mu\nu} = g_{\nu\mu},
\end{equation}
where the Greek indices $\mu,\nu$ refer to 4-dimensional spacetime coordinates.
Relaxing this rule brings about the geometry of Riemann-Cartan, which naturally includes torsion and curvature as its main geometric entities. Torsion appears naturally in the differential forms approach \cite{forms}. In this formalism, the metric tensor $g$ is written in terms of the 2-form basis $dx^{\mu}\wedge dx^{\nu}$ as $g = g_{\mu\nu} dx^{\mu}\wedge dx^{\nu}$. 

In a flat manifold, one can have a universal Cartesian frame such that $g_{\mu\nu}$ is diagonal. Although this is not possible in a more general manifold with curvature and/or torsion, one can still have it locally since the (Lorentz) tangent space in each point is flat. Therefore, it is interesting to make a  transformation between the manifold coordinates and the local Cartesian coordinates in tangent space such that the metric tensor is made (locally) diagonal. Denoting the transformation between the manifold $\theta^a$ and Lorentz $dx^{\mu}$ 1-form bases by
\begin{equation}
\theta^a = e^a_{\mu} dx^{\mu}, \label{transf}
\end{equation}
we have for the line element
\begin{equation}
ds^2 = g_{\mu\nu}dx^{\mu}dx^{\nu}=\eta_{ab} \theta^a  \theta^b, \label{diagline}
\end{equation}
where $\eta_{ab}$ is diag(-1,1,1,1) since the tangent space is flat. The transformation matrix $e^a_{\mu}$ is known as tetrad and it gives rise to the Cartan connection, as seen below.

The Cartan connection is defined as
\begin{equation}
\Gamma^{\sigma}_{\mu\nu}= e_a^{\sigma} \partial_{\nu} e^a_{\mu}, \label{Cconnection}
\end{equation}
where $e_a^{\mu}$ is such that  $e^a_{\mu}e_a^{\nu} = \delta_{\mu}^{\nu}$. This is the connection that parallel transports the tetrad field. That is, the covariant derivative
\begin{equation}
\nabla_{\nu} e^a_{\mu} = \partial_{\nu} e^a_{\mu} - \Gamma^{\rho}_{\mu\nu}e^a_{\rho} = 0.
\end{equation}
The torsion 2-form $T$ is given by the structure equation
\begin{equation}
T = d \theta + \omega \wedge \theta , \label{Tform}
\end{equation}
where $\omega$ is the 1-form spin connection. In terms of components, we have
\begin{equation}
T^{\sigma}=d \theta^{\sigma} + \omega^{\sigma}_{\nu} \wedge \theta^{\nu} = T^{\sigma}_{\mu\nu} \theta^{\mu} \wedge \theta^{\nu},
\end{equation}
where
\begin{equation}
T^{\sigma}_{\mu\nu} = \Gamma^{\sigma}_{\mu\nu} - \Gamma^{\sigma}_{\nu\mu} , \label{T}
\end{equation}
which naturally vanishes for the torsion-free Levi-Civita  connection, which is symmetric under $\mu \leftrightarrow \nu$.

Riemann-Cartan geometry applies naturally to the physics of continuum elastic media with topological defects. The defects appear as a result of breaking the translational and/or rotational symmetry of the continuum. In particular, line defects like disclinations which carry curvature but no torsion, are associated to rotational symmetry breaking. Dislocations, on the other hand, carry torsion but not curvature and are associated to translational symmetry breaking.

Consider a medium with a continuous and uniform screw dislocation distribution oriented along the z-axis.
The  metric that corresponds to a set of parallel screw dislocations, in cylindrical coordinates,  is given by Eq.  (\ref{metricdens}) :
\begin{equation}
ds^2= d\rho^2 + \rho^2 d\varphi^2 + \left( dz + \Omega \rho ^2 d\varphi \right)^2,
\label{metric}
\end{equation}
with the following density of the Burgers vectors $\Omega = bN / 2$, where $b$ is the Burgers vector and $N$ is the surface density of dislocations. This metric describes a continuous distribution of screw dislocations, with torsion uniformly distributed throughout the space.


Looking at the metric (\ref{metric}) we  choose the 1-form basis 
\begin{eqnarray}
\theta^1 & =  & d\rho \nonumber \\
\theta^2 & = &    \rho d\varphi \nonumber \\
\theta^3 & = &   dz + \Omega \rho^2 d \varphi \label{basisT}
\end{eqnarray}
such that the line element $ds^2= \delta_{ab}\theta^a \theta^b $, with $a,b = 1,2,3$. 


The torsion 2-form  is given by the structure equation (\ref{Tform})
where $\omega$ now  is zero since there is no curvature. It follows that
\begin{equation}
T = d\theta^3 = 2\Omega \,  d\rho  \wedge \rho d\varphi . \label{T2}
\end{equation}


\section{Torsion in quantum mechanics \label{tqm}}

Torsion  makes its appearance in quantum mechanics through the  axial vector 
\begin{equation}
S^{\nu} =\epsilon^{\alpha\beta\mu\nu}T_{\alpha\beta\mu}, \label{S}
\end{equation}
where $T_{\alpha\beta\mu}=g_{\alpha\sigma} T^{\sigma}_{\beta\mu} $. Since torsion couples to spin, it naturally appears in the Dirac equation.
Indeed, the Dirac operator in a  space-time with torsion can be written as \cite{shapiro}
\begin{equation}
i\hbar \frac{\partial}{\partial t} =  c \vec{\alpha}\cdot \vec{p} - \eta \vec{\alpha} \cdot \vec{S} \gamma_5 + \eta \gamma_5 S_0 + \mu c^2 \beta , \label{Dirac1}
\end{equation}
where $\eta$ is a coupling constant between torsion and matter fields and $\vec{\alpha}$, $\gamma_5$ and $\beta$ are the usual Dirac matrices. While $S_0$ is the time-axis component of the axial quadrivector, $\vec{S}$ represents its space part.  In the non-relativistic limit, this results in the low-energy Pauli-like Hamiltonian   \cite{shapiro}
\begin{equation}
H= \frac{1}{2\mu}\vec{\pi}^2 + B_0 + \vec{\sigma} \cdot \vec{Q}, \label{H}
\end{equation}
where 
\begin{equation}
\vec{\pi}=\vec{p}-\frac{\eta_1}{c}\vec{\sigma}S_0, \label{pi}
\end{equation}

\begin{equation}
\vec{Q}= \eta \vec{S} \label{Q}
\end{equation}
and
\begin{equation}
B_0 = - \frac{1}{\mu c^2}\eta^2 S_0^2 . \label{B}
\end{equation}
In Eqs. (\ref{H}) and (\ref{pi}), $\vec{\sigma}$ is the Pauli vector.

The three-dimensional version of the torsion axial vector (\ref{S}) is $S^{\nu} = \epsilon^{\alpha\beta\nu} T_{\alpha\beta}$ and its only non-null component is, from (\ref{T2}), 
\begin{equation}
S^{3} = 2 \Omega . \label{S3}
\end{equation}

It follows from Eqs. (\ref{pi})-(\ref{B}) that $\vec{\pi}=\vec{p}$, $\vec{Q}=\eta \vec{S}$,  $B_0=0$ and therefore that
\begin{equation}
H= \frac{1}{2\mu}\vec{p}\,^2  + \eta \vec{\sigma} \cdot \vec{S} , \label{Htor}
\end{equation}
from Eq. (\ref{H}).  
From Eq. (\ref{Htor}), it follows that
\begin{equation}
H=-\frac{\hbar^2}{2\mu}\nabla^2_{LB} + 2\Omega \eta \sigma^3 , \label{Hsig}
\end{equation}
where  $\nabla^2_{LB} $ stands for the Laplace-Beltrami operator, which incorporates the boundary conditions associated to the defect distribution, and $\sigma^3$ is the Pauli matrix. Since the Hamiltonian given by Eq. (\ref{Hsig}) is diagonal in the spin degree of freedom, we can write
\begin{eqnarray}
&&\left[ -\frac{\hbar^2}{2\mu}\frac{1}{\sqrt{|g|}} \frac{\partial}{\partial x^i}\left( \sqrt{|g|} g^{ij} \frac{\partial}{\partial x^j}\right) +2\Omega \eta \lambda \right] \Psi = E \Psi , \nonumber \\
\label{sch}
\end{eqnarray}
where $\lambda=\pm 1$ is the eigenvalue of the Pauli matrix $\sigma^3$.
This results in the following equation:

\begin{eqnarray}
&&\left[ \frac{1}{\rho}\frac{\partial}{\partial \rho} \left( \rho \frac{\partial}{\partial \rho} \right) + \frac{1}{\rho^2}\frac{\partial^2}{\partial \varphi^2} - 2\Omega \frac{\partial^2}{\partial z \partial \varphi} \right. + \nonumber \\
&& \left. \left( 1+\Omega^2 \rho^2 \right)\frac{\partial^2}{\partial z^2}+\frac{2\mu E}{\hbar^2}-\frac{4\mu \Omega \eta \lambda}{\hbar^2} \right] \Psi = 0 ,
\label{schrod}
\end{eqnarray}
which is Eq. (\ref{schrod1}).

{\bf Acknowledgements}: This work was supported by CAPES, CNPq and FACEPE (Brazilian agencies).

\bibliography{ref}

\begin{thebibliography}{38}%
\makeatletter
\providecommand \@ifxundefined [1]{%
 \@ifx{#1\undefined}
}%
\providecommand \@ifnum [1]{%
 \ifnum #1\expandafter \@firstoftwo
 \else \expandafter \@secondoftwo
 \fi
}%
\providecommand \@ifx [1]{%
 \ifx #1\expandafter \@firstoftwo
 \else \expandafter \@secondoftwo
 \fi
}%
\providecommand \natexlab [1]{#1}%
\providecommand \enquote  [1]{``#1''}%
\providecommand \bibnamefont  [1]{#1}%
\providecommand \bibfnamefont [1]{#1}%
\providecommand \citenamefont [1]{#1}%
\providecommand \href@noop [0]{\@secondoftwo}%
\providecommand \href [0]{\begingroup \@sanitize@url \@href}%
\providecommand \@href[1]{\@@startlink{#1}\@@href}%
\providecommand \@@href[1]{\endgroup#1\@@endlink}%
\providecommand \@sanitize@url [0]{\catcode `\\12\catcode `\$12\catcode
  `\&12\catcode `\#12\catcode `\^12\catcode `\_12\catcode `\%12\relax}%
\providecommand \@@startlink[1]{}%
\providecommand \@@endlink[0]{}%
\providecommand \url  [0]{\begingroup\@sanitize@url \@url }%
\providecommand \@url [1]{\endgroup\@href {#1}{\urlprefix }}%
\providecommand \urlprefix  [0]{URL }%
\providecommand \Eprint [0]{\href }%
\providecommand \doibase [0]{http://dx.doi.org/}%
\providecommand \selectlanguage [0]{\@gobble}%
\providecommand \bibinfo  [0]{\@secondoftwo}%
\providecommand \bibfield  [0]{\@secondoftwo}%
\providecommand \translation [1]{[#1]}%
\providecommand \BibitemOpen [0]{}%
\providecommand \bibitemStop [0]{}%
\providecommand \bibitemNoStop [0]{.\EOS\space}%
\providecommand \EOS [0]{\spacefactor3000\relax}%
\providecommand \BibitemShut  [1]{\csname bibitem#1\endcsname}%
\let\auto@bib@innerbib\@empty
\bibitem [{\citenamefont {Santos}\ \emph {et~al.}(2016)\citenamefont {Santos},
  \citenamefont {Fumeron}, \citenamefont {Berche},\ and\ \citenamefont
  {Moraes}}]{nano}%
  \BibitemOpen
  \bibfield  {author} {\bibinfo {author} {\bibfnamefont {F.}~\bibnamefont
  {Santos}}, \bibinfo {author} {\bibfnamefont {S.}~\bibnamefont {Fumeron}},
  \bibinfo {author} {\bibfnamefont {B.}~\bibnamefont {Berche}}, \ and\ \bibinfo
  {author} {\bibfnamefont {F.}~\bibnamefont {Moraes}},\ }\href@noop {}
  {\bibfield  {journal} {\bibinfo  {journal} {Nanotechnology}\ }\textbf
  {\bibinfo {volume} {27}},\ \bibinfo {pages} {135302} (\bibinfo {year}
  {2016})}\BibitemShut {NoStop}%
\bibitem [{\citenamefont {Katanaev}\ and\ \citenamefont
  {Volovich}(1992)}]{Katanaev19921}%
  \BibitemOpen
  \bibfield  {author} {\bibinfo {author} {\bibfnamefont {M.}~\bibnamefont
  {Katanaev}}\ and\ \bibinfo {author} {\bibfnamefont {I.}~\bibnamefont
  {Volovich}},\ }\href {\doibase
  http://dx.doi.org/10.1016/0003-4916(52)90040-7} {\bibfield  {journal}
  {\bibinfo  {journal} {Annals of Physics}\ }\textbf {\bibinfo {volume}
  {216}},\ \bibinfo {pages} {1 } (\bibinfo {year} {1992})}\BibitemShut
  {NoStop}%
\bibitem [{\citenamefont {Katanaev}(2005)}]{katanaev}%
  \BibitemOpen
  \bibfield  {author} {\bibinfo {author} {\bibfnamefont {M.~O.}\ \bibnamefont
  {Katanaev}},\ }\href@noop {} {\bibfield  {journal} {\bibinfo  {journal}
  {Physics-Uspekhi}\ }\textbf {\bibinfo {volume} {48}},\ \bibinfo {pages} {675}
  (\bibinfo {year} {2005})}\BibitemShut {NoStop}%
\bibitem [{\citenamefont {Moraes}(2000)}]{MORAES2000}%
  \BibitemOpen
  \bibfield  {author} {\bibinfo {author} {\bibfnamefont {F.}~\bibnamefont
  {Moraes}},\ }\href@noop {} {\bibfield  {journal} {\bibinfo  {journal}
  {{Brazilian Journal of Physics}}\ }\textbf {\bibinfo {volume} {30}},\
  \bibinfo {pages} {304 } (\bibinfo {year} {2000})}\BibitemShut {NoStop}%
\bibitem [{\citenamefont {Volterra}(1907)}]{Volterra1907}%
  \BibitemOpen
  \bibfield  {author} {\bibinfo {author} {\bibfnamefont {V.}~\bibnamefont
  {Volterra}},\ }\href@noop {} {\bibfield  {journal} {\bibinfo  {journal}
  {Annales Scientifiques de l'\'Ecole Normale Sup\'erieure}\ }\textbf {\bibinfo
  {volume} {24}},\ \bibinfo {pages} {401} (\bibinfo {year} {1907})}\BibitemShut
  {NoStop}%
\bibitem [{\citenamefont {Kondo}(1958)}]{kondo1958memoirs}%
  \BibitemOpen
  \bibfield  {author} {\bibinfo {author} {\bibfnamefont {K.}~\bibnamefont
  {Kondo}},\ }\href@noop {} {\emph {\bibinfo {title} {Memoirs of the unifying
  study of the basic problems in engineering sciences by means of geometry}}}\
  (\bibinfo  {publisher} {Gakujutsu Bunken Fukyu-Kai},\ \bibinfo {year}
  {1958})\BibitemShut {NoStop}%
\bibitem [{\citenamefont {Bilby}\ \emph {et~al.}(1955)\citenamefont {Bilby},
  \citenamefont {Bullough},\ and\ \citenamefont {Smith}}]{Bilby}%
  \BibitemOpen
  \bibfield  {author} {\bibinfo {author} {\bibfnamefont {B.~A.}\ \bibnamefont
  {Bilby}}, \bibinfo {author} {\bibfnamefont {R.}~\bibnamefont {Bullough}}, \
  and\ \bibinfo {author} {\bibfnamefont {E.}~\bibnamefont {Smith}},\ }\href
  {\doibase 10.1098/rspa.1955.0171} {\bibfield  {journal} {\bibinfo  {journal}
  {Proceedings of the Royal Society of London A: Mathematical, Physical and
  Engineering Sciences}\ }\textbf {\bibinfo {volume} {231}},\ \bibinfo {pages}
  {263} (\bibinfo {year} {1955})}\BibitemShut {NoStop}%
\bibitem [{\citenamefont {Eshelby}(1956)}]{eshelby1956continuum}%
  \BibitemOpen
  \bibfield  {author} {\bibinfo {author} {\bibfnamefont {J.}~\bibnamefont
  {Eshelby}},\ }\href@noop {} {\bibfield  {journal} {\bibinfo  {journal} {Solid
  State Physics}\ }\textbf {\bibinfo {volume} {3}},\ \bibinfo {pages} {79}
  (\bibinfo {year} {1956})}\BibitemShut {NoStop}%
\bibitem [{\citenamefont {Kr{\"o}ner}\ and\ \citenamefont
  {Rieder}(1956)}]{kroner1956kontinuumstheorie}%
  \BibitemOpen
  \bibfield  {author} {\bibinfo {author} {\bibfnamefont {E.}~\bibnamefont
  {Kr{\"o}ner}}\ and\ \bibinfo {author} {\bibfnamefont {G.}~\bibnamefont
  {Rieder}},\ }\href@noop {} {\bibfield  {journal} {\bibinfo  {journal}
  {Zeitschrift f{\"u}r Physik A Hadrons and Nuclei}\ }\textbf {\bibinfo
  {volume} {145}},\ \bibinfo {pages} {424} (\bibinfo {year}
  {1956})}\BibitemShut {NoStop}%
\bibitem [{\citenamefont {Zorawski}(1967)}]{zorawski1967theorie}%
  \BibitemOpen
  \bibfield  {author} {\bibinfo {author} {\bibfnamefont {M.}~\bibnamefont
  {Zorawski}},\ }\href@noop {} {\emph {\bibinfo {title} {Th{\'e}orie
  math{\'e}matique des dislocations}}}\ (\bibinfo  {publisher} {Dunod},\
  \bibinfo {year} {1967})\BibitemShut {NoStop}%
\bibitem [{\citenamefont {Kr{\"o}ner}(1981)}]{kroner1981continuum}%
  \BibitemOpen
  \bibfield  {author} {\bibinfo {author} {\bibfnamefont {E.}~\bibnamefont
  {Kr{\"o}ner}},\ }\href@noop {} {\emph {\bibinfo {title} {Continuum theory of
  defects in Physics of Defects (Les Houches, Session 35)}}},\ edited by\
  \bibinfo {editor} {\bibfnamefont {R.}~\bibnamefont {Balian}} \emph {et~al.},\
  Vol.~\bibinfo {volume} {35}\ (\bibinfo  {publisher} {North-Holland,
  Amsterdam},\ \bibinfo {year} {1981})\ pp.\ \bibinfo {pages}
  {215--315}\BibitemShut {NoStop}%
\bibitem [{\citenamefont {Maurel}\ \emph {et~al.}(2008)\citenamefont {Maurel},
  \citenamefont {Pagneux}, \citenamefont {Barra},\ and\ \citenamefont
  {Lund}}]{maurel2008interaction}%
  \BibitemOpen
  \bibfield  {author} {\bibinfo {author} {\bibfnamefont {A.}~\bibnamefont
  {Maurel}}, \bibinfo {author} {\bibfnamefont {V.}~\bibnamefont {Pagneux}},
  \bibinfo {author} {\bibfnamefont {F.}~\bibnamefont {Barra}}, \ and\ \bibinfo
  {author} {\bibfnamefont {F.}~\bibnamefont {Lund}},\ }\href@noop {} {\bibfield
   {journal} {\bibinfo  {journal} {The Journal of the Acoustical Society of
  America}\ }\textbf {\bibinfo {volume} {123}},\ \bibinfo {pages} {3408}
  (\bibinfo {year} {2008})}\BibitemShut {NoStop}%
\bibitem [{\citenamefont {Woltersdorf}\ and\ \citenamefont
  {Heinrich}(2004)}]{woltersdorf2004two}%
  \BibitemOpen
  \bibfield  {author} {\bibinfo {author} {\bibfnamefont {G.}~\bibnamefont
  {Woltersdorf}}\ and\ \bibinfo {author} {\bibfnamefont {B.}~\bibnamefont
  {Heinrich}},\ }\href@noop {} {\bibfield  {journal} {\bibinfo  {journal}
  {Physical Review B}\ }\textbf {\bibinfo {volume} {69}},\ \bibinfo {pages}
  {184417} (\bibinfo {year} {2004})}\BibitemShut {NoStop}%
\bibitem [{\citenamefont {Jaszek}(2001)}]{Jaszek2001}%
  \BibitemOpen
  \bibfield  {author} {\bibinfo {author} {\bibfnamefont {R.}~\bibnamefont
  {Jaszek}},\ }\href {\doibase 10.1023/A:1011228626077} {\bibfield  {journal}
  {\bibinfo  {journal} {Journal of Materials Science: Materials in
  Electronics}\ }\textbf {\bibinfo {volume} {12}},\ \bibinfo {pages} {1}
  (\bibinfo {year} {2001})}\BibitemShut {NoStop}%
\bibitem [{\citenamefont {Ran}\ \emph {et~al.}(2009)\citenamefont {Ran},
  \citenamefont {Zhang},\ and\ \citenamefont {Vishwanath}}]{vish}%
  \BibitemOpen
  \bibfield  {author} {\bibinfo {author} {\bibfnamefont {Y.}~\bibnamefont
  {Ran}}, \bibinfo {author} {\bibfnamefont {Y.}~\bibnamefont {Zhang}}, \ and\
  \bibinfo {author} {\bibfnamefont {A.}~\bibnamefont {Vishwanath}},\
  }\href@noop {} {\bibfield  {journal} {\bibinfo  {journal} {Nature Physics}\
  }\textbf {\bibinfo {volume} {5}},\ \bibinfo {pages} {298} (\bibinfo {year}
  {2009})}\BibitemShut {NoStop}%
\bibitem [{\citenamefont {Tretiakov}\ \emph {et~al.}(2010)\citenamefont
  {Tretiakov}, \citenamefont {Abanov}, \citenamefont {Murakami},\ and\
  \citenamefont {Sinova}}]{jairo}%
  \BibitemOpen
  \bibfield  {author} {\bibinfo {author} {\bibfnamefont {O.~A.}\ \bibnamefont
  {Tretiakov}}, \bibinfo {author} {\bibfnamefont {A.}~\bibnamefont {Abanov}},
  \bibinfo {author} {\bibfnamefont {S.}~\bibnamefont {Murakami}}, \ and\
  \bibinfo {author} {\bibfnamefont {J.}~\bibnamefont {Sinova}},\ }\href
  {\doibase http://dx.doi.org/10.1063/1.3481382} {\bibfield  {journal}
  {\bibinfo  {journal} {Applied Physics Letters}\ }\textbf {\bibinfo {volume}
  {97}},\ \bibinfo {eid} {073108} (\bibinfo {year} {2010}),\
  http://dx.doi.org/10.1063/1.3481382}\BibitemShut {NoStop}%
\bibitem [{\citenamefont {Bakke}\ and\ \citenamefont {Moraes}(2012)}]{knut}%
  \BibitemOpen
  \bibfield  {author} {\bibinfo {author} {\bibfnamefont {K.}~\bibnamefont
  {Bakke}}\ and\ \bibinfo {author} {\bibfnamefont {F.}~\bibnamefont {Moraes}},\
  }\href {\doibase http://dx.doi.org/10.1016/j.physleta.2012.09.006} {\bibfield
   {journal} {\bibinfo  {journal} {Physics Letters A}\ }\textbf {\bibinfo
  {volume} {376}},\ \bibinfo {pages} {2838 } (\bibinfo {year}
  {2012})}\BibitemShut {NoStop}%
\bibitem [{\citenamefont {Sumiyoshi}\ and\ \citenamefont
  {Fujimoto}(2016)}]{Sumiyoshi}%
  \BibitemOpen
  \bibfield  {author} {\bibinfo {author} {\bibfnamefont {H.}~\bibnamefont
  {Sumiyoshi}}\ and\ \bibinfo {author} {\bibfnamefont {S.}~\bibnamefont
  {Fujimoto}},\ }\href {\doibase 10.1103/PhysRevLett.116.166601} {\bibfield
  {journal} {\bibinfo  {journal} {Phys. Rev. Lett.}\ }\textbf {\bibinfo
  {volume} {116}},\ \bibinfo {pages} {166601} (\bibinfo {year}
  {2016})}\BibitemShut {NoStop}%
\bibitem [{\citenamefont {Kawamura}(1978)}]{Kawamura1978}%
  \BibitemOpen
  \bibfield  {author} {\bibinfo {author} {\bibfnamefont {K.}~\bibnamefont
  {Kawamura}},\ }\href {\doibase 10.1007/BF01313193} {\bibfield  {journal}
  {\bibinfo  {journal} {Zeitschrift f{\"u}r Physik B Condensed Matter}\
  }\textbf {\bibinfo {volume} {29}},\ \bibinfo {pages} {101} (\bibinfo {year}
  {1978})}\BibitemShut {NoStop}%
\bibitem [{\citenamefont {Bausch}\ \emph {et~al.}(1999)\citenamefont {Bausch},
  \citenamefont {Schmitz},\ and\ \citenamefont {Turski}}]{PhysRevB.59.13491}%
  \BibitemOpen
  \bibfield  {author} {\bibinfo {author} {\bibfnamefont {R.}~\bibnamefont
  {Bausch}}, \bibinfo {author} {\bibfnamefont {R.}~\bibnamefont {Schmitz}}, \
  and\ \bibinfo {author} {\bibfnamefont {L.~A.}\ \bibnamefont {Turski}},\
  }\href {\doibase 10.1103/PhysRevB.59.13491} {\bibfield  {journal} {\bibinfo
  {journal} {Phys. Rev. B}\ }\textbf {\bibinfo {volume} {59}},\ \bibinfo
  {pages} {13491} (\bibinfo {year} {1999})}\BibitemShut {NoStop}%
\bibitem [{\citenamefont {Furtado}\ \emph {et~al.}(2001)\citenamefont
  {Furtado}, \citenamefont {Bezerra},\ and\ \citenamefont
  {Moraes}}]{Furtado2001160}%
  \BibitemOpen
  \bibfield  {author} {\bibinfo {author} {\bibfnamefont {C.}~\bibnamefont
  {Furtado}}, \bibinfo {author} {\bibfnamefont {V.}~\bibnamefont {Bezerra}}, \
  and\ \bibinfo {author} {\bibfnamefont {F.}~\bibnamefont {Moraes}},\ }\href
  {\doibase http://dx.doi.org/10.1016/S0375-9601(01)00615-6} {\bibfield
  {journal} {\bibinfo  {journal} {Physics Letters A}\ }\textbf {\bibinfo
  {volume} {289}},\ \bibinfo {pages} {160 } (\bibinfo {year}
  {2001})}\BibitemShut {NoStop}%
\bibitem [{\citenamefont {de~Lima}\ \emph {et~al.}(2013)\citenamefont
  {de~Lima}, \citenamefont {Poux}, \citenamefont {Assafr{\~a}o},\ and\
  \citenamefont {Filgueiras}}]{clev}%
  \BibitemOpen
  \bibfield  {author} {\bibinfo {author} {\bibfnamefont {G.~A.}\ \bibnamefont
  {de~Lima}}, \bibinfo {author} {\bibfnamefont {A.}~\bibnamefont {Poux}},
  \bibinfo {author} {\bibfnamefont {D.}~\bibnamefont {Assafr{\~a}o}}, \ and\
  \bibinfo {author} {\bibfnamefont {C.}~\bibnamefont {Filgueiras}},\ }\href
  {\doibase 10.1140/epjb/e2013-40160-x} {\bibfield  {journal} {\bibinfo
  {journal} {The European Physical Journal B}\ }\textbf {\bibinfo {volume}
  {86}},\ \bibinfo {pages} {1} (\bibinfo {year} {2013})}\BibitemShut {NoStop}%
\bibitem [{\citenamefont {Taira}\ and\ \citenamefont {Shima}(2014)}]{Taira}%
  \BibitemOpen
  \bibfield  {author} {\bibinfo {author} {\bibfnamefont {H.}~\bibnamefont
  {Taira}}\ and\ \bibinfo {author} {\bibfnamefont {H.}~\bibnamefont {Shima}},\
  }\href {\doibase http://dx.doi.org/10.1016/j.ssc.2013.10.002} {\bibfield
  {journal} {\bibinfo  {journal} {Solid State Communications}\ }\textbf
  {\bibinfo {volume} {177}},\ \bibinfo {pages} {61 } (\bibinfo {year}
  {2014})}\BibitemShut {NoStop}%
\bibitem [{\citenamefont {\ifmmode \check{Z}\else
  \v{Z}\fi{}uti\ifmmode~\acute{c}\else \'{c}\fi{}}\ \emph
  {et~al.}(2004)\citenamefont {\ifmmode \check{Z}\else
  \v{Z}\fi{}uti\ifmmode~\acute{c}\else \'{c}\fi{}}, \citenamefont {Fabian},\
  and\ \citenamefont {Das~Sarma}}]{RevModPhys.76.323}%
  \BibitemOpen
  \bibfield  {author} {\bibinfo {author} {\bibfnamefont {I.}~\bibnamefont
  {\ifmmode \check{Z}\else \v{Z}\fi{}uti\ifmmode~\acute{c}\else \'{c}\fi{}}},
  \bibinfo {author} {\bibfnamefont {J.}~\bibnamefont {Fabian}}, \ and\ \bibinfo
  {author} {\bibfnamefont {S.}~\bibnamefont {Das~Sarma}},\ }\href {\doibase
  10.1103/RevModPhys.76.323} {\bibfield  {journal} {\bibinfo  {journal} {Rev.
  Mod. Phys.}\ }\textbf {\bibinfo {volume} {76}},\ \bibinfo {pages} {323}
  (\bibinfo {year} {2004})}\BibitemShut {NoStop}%
\bibitem [{\citenamefont {Netto}\ and\ \citenamefont
  {Furtado}(2008)}]{elastic}%
  \BibitemOpen
  \bibfield  {author} {\bibinfo {author} {\bibfnamefont {A.~L.~S.}\
  \bibnamefont {Netto}}\ and\ \bibinfo {author} {\bibfnamefont
  {C.}~\bibnamefont {Furtado}},\ }\href@noop {} {\bibfield  {journal} {\bibinfo
   {journal} {Journal of Physics: Condensed Matter}\ }\textbf {\bibinfo
  {volume} {20}},\ \bibinfo {pages} {125209} (\bibinfo {year}
  {2008})}\BibitemShut {NoStop}%
\bibitem [{\citenamefont {Eringen}\ and\ \citenamefont
  {Maugin}(1990)}]{Eringen}%
  \BibitemOpen
  \bibfield  {author} {\bibinfo {author} {\bibfnamefont {A.~C.}\ \bibnamefont
  {Eringen}}\ and\ \bibinfo {author} {\bibfnamefont {G.~A.}\ \bibnamefont
  {Maugin}},\ }\href@noop {} {\emph {\bibinfo {title} {Electrodynamics of
  Continua I}}},\ \bibinfo {edition} {1st}\ ed.\ (\bibinfo  {publisher}
  {Springer-Verlag New York},\ \bibinfo {year} {1990})\ \bibinfo {note}
  {chapter 7}\BibitemShut {NoStop}%
\bibitem [{\citenamefont {Seeger}\ \emph {et~al.}(1964)\citenamefont {Seeger},
  \citenamefont {Kronmüller}, \citenamefont {Rieger},\ and\ \citenamefont
  {Träuble}}]{Seeger}%
  \BibitemOpen
  \bibfield  {author} {\bibinfo {author} {\bibfnamefont {A.}~\bibnamefont
  {Seeger}}, \bibinfo {author} {\bibfnamefont {H.}~\bibnamefont {Kronmüller}},
  \bibinfo {author} {\bibfnamefont {H.}~\bibnamefont {Rieger}}, \ and\ \bibinfo
  {author} {\bibfnamefont {H.}~\bibnamefont {Träuble}},\ }\href {\doibase
  http://dx.doi.org/10.1063/1.1713460} {\bibfield  {journal} {\bibinfo
  {journal} {Journal of Applied Physics}\ }\textbf {\bibinfo {volume} {35}},\
  \bibinfo {pages} {740} (\bibinfo {year} {1964})}\BibitemShut {NoStop}%
\bibitem [{\citenamefont {Berbil-Bautista}\ \emph {et~al.}(2009)\citenamefont
  {Berbil-Bautista}, \citenamefont {Krause}, \citenamefont {Bode},
  \citenamefont {Bad\'{\i}a-Maj\'os}, \citenamefont {de~la Fuente},
  \citenamefont {Wiesendanger},\ and\ \citenamefont {Arnaudas}}]{Bode}%
  \BibitemOpen
  \bibfield  {author} {\bibinfo {author} {\bibfnamefont {L.}~\bibnamefont
  {Berbil-Bautista}}, \bibinfo {author} {\bibfnamefont {S.}~\bibnamefont
  {Krause}}, \bibinfo {author} {\bibfnamefont {M.}~\bibnamefont {Bode}},
  \bibinfo {author} {\bibfnamefont {A.}~\bibnamefont {Bad\'{\i}a-Maj\'os}},
  \bibinfo {author} {\bibfnamefont {C.}~\bibnamefont {de~la Fuente}}, \bibinfo
  {author} {\bibfnamefont {R.}~\bibnamefont {Wiesendanger}}, \ and\ \bibinfo
  {author} {\bibfnamefont {J.~I.}\ \bibnamefont {Arnaudas}},\ }\href {\doibase
  10.1103/PhysRevB.80.241408} {\bibfield  {journal} {\bibinfo  {journal} {Phys.
  Rev. B}\ }\textbf {\bibinfo {volume} {80}},\ \bibinfo {pages} {241408}
  (\bibinfo {year} {2009})}\BibitemShut {NoStop}%
\bibitem [{\citenamefont {Shapiro}(2002)}]{shapiro}%
  \BibitemOpen
  \bibfield  {author} {\bibinfo {author} {\bibfnamefont {I.}~\bibnamefont
  {Shapiro}},\ }\href {\doibase
  http://dx.doi.org/10.1016/S0370-1573(01)00030-8} {\bibfield  {journal}
  {\bibinfo  {journal} {Physics Reports}\ }\textbf {\bibinfo {volume} {357}},\
  \bibinfo {pages} {113 } (\bibinfo {year} {2002})}\BibitemShut {NoStop}%
\bibitem [{\citenamefont {Rollett}\ \emph {et~al.}(2004)\citenamefont
  {Rollett}, \citenamefont {Humphreys}, \citenamefont {Rohrer},\ and\
  \citenamefont {Hatherly}}]{rollett2004recrystallization}%
  \BibitemOpen
  \bibfield  {author} {\bibinfo {author} {\bibfnamefont {A.}~\bibnamefont
  {Rollett}}, \bibinfo {author} {\bibfnamefont {F.}~\bibnamefont {Humphreys}},
  \bibinfo {author} {\bibfnamefont {G.~S.}\ \bibnamefont {Rohrer}}, \ and\
  \bibinfo {author} {\bibfnamefont {M.}~\bibnamefont {Hatherly}},\ }\href@noop
  {} {\emph {\bibinfo {title} {Recrystallization and related annealing
  phenomena}}}\ (\bibinfo  {publisher} {Elsevier},\ \bibinfo {year}
  {2004})\BibitemShut {NoStop}%
\bibitem [{\citenamefont {Symon}(1971)}]{symon1971mechanics}%
  \BibitemOpen
  \bibfield  {author} {\bibinfo {author} {\bibfnamefont {K.~R.}\ \bibnamefont
  {Symon}},\ }\href@noop {} {\emph {\bibinfo {title} {Mechanics}}}\ (\bibinfo
  {publisher} {Addison-Wesley},\ \bibinfo {year} {1971})\BibitemShut {NoStop}%
\bibitem [{\citenamefont {Bagrov}\ \emph {et~al.}(1992)\citenamefont {Bagrov},
  \citenamefont {Buchbinder},\ and\ \citenamefont {I.L.}}]{shapiro2}%
  \BibitemOpen
  \bibfield  {author} {\bibinfo {author} {\bibfnamefont {V.}~\bibnamefont
  {Bagrov}}, \bibinfo {author} {\bibfnamefont {I.}~\bibnamefont {Buchbinder}},
  \ and\ \bibinfo {author} {\bibfnamefont {S.}~\bibnamefont {I.L.}},\
  }\href@noop {} {\bibfield  {journal} {\bibinfo  {journal} {Izv. VUZov,
  Fisica-Sov. J. Phys.}\ }\textbf {\bibinfo {volume} {35}},\ \bibinfo {pages}
  {5} (\bibinfo {year} {1992})}\BibitemShut {NoStop}%
\bibitem [{\citenamefont {Ryder}\ and\ \citenamefont
  {Shapiro}(1998)}]{Ryder199821}%
  \BibitemOpen
  \bibfield  {author} {\bibinfo {author} {\bibfnamefont {L.~H.}\ \bibnamefont
  {Ryder}}\ and\ \bibinfo {author} {\bibfnamefont {I.~L.}\ \bibnamefont
  {Shapiro}},\ }\href {\doibase
  http://dx.doi.org/10.1016/S0375-9601(98)00503-9} {\bibfield  {journal}
  {\bibinfo  {journal} {Physics Letters A}\ }\textbf {\bibinfo {volume}
  {247}},\ \bibinfo {pages} {21 } (\bibinfo {year} {1998})}\BibitemShut
  {NoStop}%
\bibitem [{\citenamefont {Kato}\ \emph {et~al.}(2004)\citenamefont {Kato},
  \citenamefont {Myers}, \citenamefont {Gossard},\ and\ \citenamefont
  {Awschalom}}]{kato2004coherent}%
  \BibitemOpen
  \bibfield  {author} {\bibinfo {author} {\bibfnamefont {Y.}~\bibnamefont
  {Kato}}, \bibinfo {author} {\bibfnamefont {R.}~\bibnamefont {Myers}},
  \bibinfo {author} {\bibfnamefont {A.}~\bibnamefont {Gossard}}, \ and\
  \bibinfo {author} {\bibfnamefont {D.}~\bibnamefont {Awschalom}},\ }\href@noop
  {} {\bibfield  {journal} {\bibinfo  {journal} {Nature}\ }\textbf {\bibinfo
  {volume} {427}},\ \bibinfo {pages} {50} (\bibinfo {year} {2004})}\BibitemShut
  {NoStop}%
\bibitem [{\citenamefont {Chantis}\ \emph {et~al.}(2008)\citenamefont
  {Chantis}, \citenamefont {Cardona}, \citenamefont {Christensen},
  \citenamefont {Smith}, \citenamefont {Van~Schilfgaarde}, \citenamefont
  {Kotani}, \citenamefont {Svane},\ and\ \citenamefont
  {Albers}}]{chantis2008strain}%
  \BibitemOpen
  \bibfield  {author} {\bibinfo {author} {\bibfnamefont {A.~N.}\ \bibnamefont
  {Chantis}}, \bibinfo {author} {\bibfnamefont {M.}~\bibnamefont {Cardona}},
  \bibinfo {author} {\bibfnamefont {N.~E.}\ \bibnamefont {Christensen}},
  \bibinfo {author} {\bibfnamefont {D.~L.}\ \bibnamefont {Smith}}, \bibinfo
  {author} {\bibfnamefont {M.}~\bibnamefont {Van~Schilfgaarde}}, \bibinfo
  {author} {\bibfnamefont {T.}~\bibnamefont {Kotani}}, \bibinfo {author}
  {\bibfnamefont {A.}~\bibnamefont {Svane}}, \ and\ \bibinfo {author}
  {\bibfnamefont {R.~C.}\ \bibnamefont {Albers}},\ }\href@noop {} {\bibfield
  {journal} {\bibinfo  {journal} {Physical Review B}\ }\textbf {\bibinfo
  {volume} {78}},\ \bibinfo {pages} {075208} (\bibinfo {year}
  {2008})}\BibitemShut {NoStop}%
\bibitem [{\citenamefont {Jr}(1996)}]{esr}%
  \BibitemOpen
  \bibfield  {author} {\bibinfo {author} {\bibfnamefont {C.~P.~P.}\
  \bibnamefont {Jr}},\ }\href@noop {} {\emph {\bibinfo {title} {Electron Spin
  Resonace: A Comprehensive Treatise on Experimental Techniques}}},\ \bibinfo
  {edition} {2nd}\ ed.\ (\bibinfo  {publisher} {Dover, New York},\ \bibinfo
  {year} {1996})\BibitemShut {NoStop}%
\bibitem [{\citenamefont {Li}\ and\ \citenamefont {Li}(2008)}]{li2008strain}%
  \BibitemOpen
  \bibfield  {author} {\bibinfo {author} {\bibfnamefont {Y.}~\bibnamefont
  {Li}}\ and\ \bibinfo {author} {\bibfnamefont {Y.-Q.}\ \bibnamefont {Li}},\
  }\href@noop {} {\bibfield  {journal} {\bibinfo  {journal} {The European
  Physical Journal B}\ }\textbf {\bibinfo {volume} {63}},\ \bibinfo {pages}
  {493} (\bibinfo {year} {2008})}\BibitemShut {NoStop}%
\bibitem [{\citenamefont {do~Carmo}(1994)}]{forms}%
  \BibitemOpen
  \bibfield  {author} {\bibinfo {author} {\bibfnamefont {M.~P.}\ \bibnamefont
  {do~Carmo}},\ }\href@noop {} {\emph {\bibinfo {title} {Differential Forms and
  Applications}}},\ \bibinfo {edition} {1st}\ ed.\ (\bibinfo  {publisher}
  {Springer-Verlag Berlin},\ \bibinfo {year} {1994})\BibitemShut {NoStop}%
\end{thebibliography}%

\end{document}